\pdfoutput=1

\documentclass[11pt,twoside,a4paper,cmspaper,final,collab]{cms-tdr}
\def\svnVersion{285778}\def\cmsCernNo{2013-037}\def\cmsCernDate{\today}\def\cmsMessage{Published in Physics Letters B as \href{http://dx.doi.org/10.1106/j.physletb.2015.03.048}{\doi{10.1106/j.physletb.2015.03.048.}}}

\begin{document}\cmsNoteHeader{HIG-13-007}

\hyphenation{had-ron-i-za-tion}
\hyphenation{cal-or-i-me-ter}
\hyphenation{de-vices}
\RCS$Revision: 278149 $
\RCS$HeadURL: svn+ssh://svn.cern.ch/reps/tdr2/papers/HIG-13-007/trunk/HIG-13-007.tex $
\RCS$Id: HIG-13-007.tex 278149 2015-02-19 20:52:20Z alverson $

\hyphenation{ATLAS}
\hyphenation{had-ron-i-za-tion}
\hyphenation{cal-or-i-me-ter}
\hyphenation{de-vices}

\newcommand{\mc}[1]{\multicolumn{1}{c}{#1}}
\newcommand{\mcd}{\mc{---}}

\newcommand{\mh}{\ensuremath{m_\PH}\xspace}
\newcommand{\mz}{\ensuremath{m_\cPZ}\xspace}
\newcommand{\ptmumu}{\ensuremath{\pt^{\Pgm\Pgm}}\xspace}
\newcommand{\ptll}{\ensuremath{\pt^{\ell\ell}}\xspace}
\newcommand{\Mmumu}{\ensuremath{m_{\Pgm\Pgm}}\xspace}
\newcommand{\Mee}{\ensuremath{m_{\Pe\Pe}}\xspace}
\newcommand{\Mll}{\ensuremath{m_{\ell\ell}}\xspace}
\newcommand{\Mjj}{\ensuremath{m_{\mathrm{jj}}}\xspace}
\newcommand{\etajj}{\ensuremath{\eta(\mathrm{jj})}\xspace}
\newcommand{\ptmiss}{\ensuremath{\pt^{\text{miss}}}}

\newcommand{\mumu}{\ensuremath{\Pgmp\Pgmm}\xspace}
\newcommand{\ee}{\ensuremath{\Pep\Pem}\xspace}
\newcommand{\tautau}{\ensuremath{\tau^{+}\tau^{-}}\xspace}
\newcommand{\gamgam}{\ensuremath{\gamma\gamma}}
\newcommand{\zz}{\ensuremath{\cPZ\cPZ}\xspace}
\newcommand{\ellell}{\ensuremath{\ell^+\ell^-}\xspace}

\newcommand{\hmm}{\ensuremath{\PH\to\mumu}\xspace}
\newcommand{\hee}{\ensuremath{\PH\to\ee}\xspace}
\newcommand{\hll}{\ensuremath{\PH\to\ellell}\xspace}
\newcommand{\hgamgam}{\ensuremath{\PH\to\gamgam}\xspace}
\newcommand{\htautau}{\ensuremath{\PH\to\tautau}\xspace}
\newcommand{\hzz}{\ensuremath{\PH\to\zz}\xspace}

\newcommand{\BF}{\ensuremath{\mathcal{B}}\xspace}

\newcolumntype{d}[1]{D{.}{.}{#1}}
\ifthenelse{\boolean{cms@external}}{\providecommand{\cmsLeft}{top\xspace}}{\providecommand{\cmsLeft}{left\xspace}}
\ifthenelse{\boolean{cms@external}}{\providecommand{\cmsRight}{bottom\xspace}}{\providecommand{\cmsRight}{right\xspace}}

\cmsNoteHeader{HIG-13-007}

\title{Search for a standard model-like Higgs boson in the $\Pgmp\Pgmm$ and $\Pep\Pem$ decay channels at the LHC}

\author{The CMS Collaboration}
\date{\today}

\abstract{
A search is presented for a standard model-like Higgs boson decaying to the \mumu or
\ee final states based on
proton-proton collisions recorded by the CMS experiment at the CERN LHC. The data
correspond to integrated luminosities of 5.0\fbinv at a
centre-of-mass energy of 7\TeV and
19.7\fbinv at 8\TeV for the \mumu search,
and of 19.7\fbinv at 8\TeV for the \ee search.
Upper limits on the production cross section times branching fraction at the 95\% confidence level
are reported for Higgs boson masses in the range  from 120 to 150\GeV.
For a Higgs boson with a mass of 125\GeV decaying to \mumu, the observed
(expected) upper limit on the production rate is found to be 7.4 ($6.5^{+2.8}_{-1.9}$) times
the standard model value. This corresponds to an upper limit on the branching fraction of 0.0016.
Similarly, for
\ee, an upper limit of 0.0019 is placed on the branching fraction,
which is ${\approx}3.7\times10^5$ times the standard model value.
These results, together with recent evidence of the 125\GeV boson coupling to
$\tau$-leptons with a larger branching fraction consistent with the standard model,
confirm that the leptonic couplings of the new boson are not flavour-universal.
}

 \hypersetup{%
pdfauthor={CMS Collaboration},%
pdftitle={Search for a standard model-like Higgs boson in the mu+ mu- and e+ e- decay channels at the LHC},%
pdfsubject={CMS},%
pdfkeywords={CMS, physics, Higgs, lepton, muon, electron, dilepton, dimuon, dielectron, Higgs boson}}

\maketitle 

 \section{Introduction}

After the discovery of a particle with a mass near 125\GeV~\cite{ATLASDiscovery,CMSDiscovery,CMSDiscoveryLong} and properties
in agreement, within current experimental uncertainties, with those expected of the standard model (SM) Higgs boson,
the next critical question is to understand in greater detail the nature of the newly discovered particle.  Answering
this question with a reasonable confidence requires measurements of its properties and production rates into final
states both allowed and disallowed by the SM.  Beyond the standard model (BSM) scenarios may contain
additional Higgs bosons, so searches for these additional states constitute another test
of the SM~\cite{2HDM}.
For a Higgs boson mass, \mh, of 125\GeV, the SM prediction for the Higgs to
\mumu branching fraction, $\BF(\hmm)$,
is among the smallest accessible at the CERN LHC,
$2.2\times 10^{-4}$~\cite{Denner_2011mq},
while the SM prediction for \BF(\hee) of approximately $5\times10^{-9}$
is inaccessible at the LHC.
Experimentally, however, \hmm and \hee are the cleanest of the fermionic decays.
The clean final states allow a better sensitivity, in terms of
cross section, $\sigma$, times branching fraction, \BF,
than \htautau.
This means that searches for \hmm and \hee, combined with recent
strong evidence for decays of the new boson to \tautau~\cite{cmsHtautau,atlasHtautau},
may be used to test if the coupling of the new boson to leptons is flavour-universal or proportional
to the lepton mass, as predicted by the SM~\cite{Weinberg:1967tq}.
In addition, a measurement of the \hmm decay
probes the Yukawa coupling of the Higgs boson to
second-generation fermions, an important input in understanding the
mechanism of electroweak symmetry breaking in the SM~\cite{Plehn:2001qg,Han:2002gp}.
Deviations from the SM expectation could also be a sign of BSM physics~\cite{Vignaroli:2009vt,newRatio}.
A previous LHC search for SM \hmm has been performed by the ATLAS Collaboration and
placed a 95\% confidence level~(CL) upper limit of 7.0 times the rate
expected from the SM at 125.5\GeV~\cite{atlasSM}.
The ATLAS Collaboration has also performed a search for BSM \hmm decays
within the context of the minimal supersymmetric standard model~\cite{atlasMSSM}.

This paper reports on a search for a SM-like Higgs boson decaying to either a pair of muons
or electrons (\hll)
in proton-proton collisions recorded by the CMS experiment at the LHC.
The \hmm search is performed on data corresponding to integrated luminosities of
$5.0\pm0.1$\fbinv at a centre-of-mass
energy of 7\TeV and $19.7\pm0.5$\fbinv at 8\TeV,
while the \hee search is only performed on the 8\TeV data.
Results are presented for Higgs boson masses between 120 and 150\GeV.
For $\mh=125\GeV$, the SM predicts 19\,(95) \hmm events
at 7\TeV\,(8\TeV), and ${\approx}2\times10^{-3}$ \hee events
at 8\TeV~\cite{deFlorian_2012yg,LHCHXSWG1,LHCHXSWG2,LHCHXSWG3}.

The \hll resonance is sought as a peak in the dilepton mass spectrum, \Mll,
on top of a smoothly falling
background dominated by contributions from Drell--Yan production,
$\ttbar$ production, and vector boson pair-production processes.
Signal acceptance and selection efficiency are estimated using Monte Carlo (MC) simulations, while
the background is estimated by fitting the observed \Mll spectrum in data, assuming a smooth functional form.

Near $\mh=125\GeV$, the SM predicts a Higgs boson decay width much narrower
than the dilepton invariant mass resolution of the CMS experiment.
For $\mh=125\GeV$,
the SM predicts the Higgs boson decay width to be 4.2\MeV~\cite{LHCHXSWG1},
and experimental results indirectly constrain the width to be ${<}22\MeV$
at the 95\% CL, subject to various assumptions~\cite{cmsHiggsWidthIndirect,Caola:2013yja}.
The experimental resolution depends on the angle of each reconstructed
lepton relative to the beam axis.
For dimuons, the full width at half maximum (FWHM) of the signal peak ranges
from 3.9 to 6.2\GeV (for muons with $\abs{\eta}<2.1$),
while for electrons it ranges from 4.0 to 7.2\GeV
(for electrons with $\abs{\eta}<1.44$ or $1.57<\abs{\eta}<2.5$).

The sensitivity of this analysis is increased through an extensive categorization of the events, using
kinematic variables to isolate regions with a large signal over background (S/B)
ratio from regions with smaller S/B ratios.
Separate categories are optimized for the dominant Higgs boson
production mode, gluon-fusion (GF), and the sub-dominant production
mode, vector boson fusion (VBF).  Higgs boson production in association
with a vector boson (VH), while not optimized for, is taken into
account in the \hmm analysis.
The SM predicts Higgs boson production to be
87.2\%~GF, 7.1\%~VBF, and 5.1\%~VH for $\mh=125\GeV$ at 8\TeV~\cite{LHCHXSWG3}.
In addition to \Mll, the most powerful variables for discriminating between the Higgs boson
signal and the Drell--Yan and $\ttbar$ backgrounds are
the jet multiplicity, the dilepton transverse-momentum (\ptll), and
the invariant mass of the two largest transverse-momentum jets (\Mjj).
The gluon-gluon initial state of GF production tends to lead to more
jet radiation than the quark-antiquark initial state of Drell--Yan production,
leading to larger \ptll and jet multiplicity.  Similarly,
VBF production involves a pair of forward-backward jets
with a large \Mjj compared to
Drell--Yan plus two-jet or $\ttbar$ production.
Events are further categorized by their \Mll resolution
and the kinematics of the jets and leptons.

This paper is organized as follows.
Section~\ref{sec:cmsDet} introduces the CMS detector and event reconstruction,
Section~\ref{sec:evtSel} describes the \hmm event selection,
Section~\ref{sec:selEff} the \hmm selection efficiency,
Section~\ref{sec:systUnc} details the systematic uncertainties included in the \hmm analysis,
Section~\ref{sec:results} presents the results of the \hmm search,
Section~\ref{sec:hee} describes the \hee search,
and Section~\ref{sec:summary} provides a summary.

 \section{CMS detector and event reconstruction}
\label{sec:cmsDet}

The central feature of the CMS apparatus is
a superconducting solenoid of 6\unit{m} internal diameter,
providing a magnetic field of 3.8\unit{T}. Within the
superconducting solenoid volume are a silicon pixel and strip
tracker, a lead tungstate crystal electromagnetic calorimeter (ECAL),
and a brass/scintillator hadron calorimeter (HCAL), each composed
of a barrel and two endcap sections. Muons are measured in
gas-ionization detectors embedded in the steel flux-return yoke
outside the solenoid. Extensive forward calorimetry complements the
coverage provided by the barrel and endcap detectors.

The first level of the CMS trigger system, composed of custom hardware processors,
uses information from the calorimeters and muon detectors to select the most
interesting events in a fixed time interval of less than 4\mus. The
high level trigger processor farm further decreases the event rate from
at most 100\unit{kHz} to less than 1\unit{kHz}, before data storage.
A more detailed description of the detector as well as the definition of the
coordinate system and relevant kinematic variables can be found in Ref.~\cite{cmsDet}.

The CMS offline event reconstruction creates a global event description by combining information
from all subdetectors. This combined information then leads to a list of particle-flow (PF)
objects~\cite{CMS-PAS-PFT-09-001,CMS-PAS-PFT-10-001}: candidate muons, electrons, photons, and hadrons.
By combining information from all subdetectors, particle identification and
energy estimation performance are improved.
In addition, double counting subdetector energy deposits when reconstructing different particle types is
eliminated.

Due to the high instantaneous luminosity of the LHC, many proton-proton interactions occur
in each bunch crossing.  An average of 9 and 21 interactions occur in each bunch crossing for
the 7 and 8\TeV data samples, respectively.  Most interactions produce particles with relatively low
transverse-momentum (\pt), compared to the particles produced in an \hll
signal event.  These interactions are termed ``pileup'', and can interfere with the reconstruction
of the high-\pt interaction, whose vertex is identified
as the vertex with the largest scalar sum of the squared transverse momenta of the tracks associated with it.
All charged PF objects with tracks coming from another vertex are then removed.

Hadronic jets are clustered from reconstructed PF objects with the infrared- and
collinear-safe anti-\kt algorithm~\cite{antikt,fastjet}, operated with a size parameter of 0.5.
 The jet momentum is determined as the vectorial sum of the momenta of all PF objects in the jet,
and is found in the simulation to be within 5\% to 10\% of the true momentum over
the whole \pt spectrum of interest and detector acceptance. An offset correction is applied to
take into account the extra neutral energy clustered in jets due to pileup.
Jet energy corrections are derived
from the simulation, and are confirmed
by in-situ measurements of the energy balance in
dijet, photon plus jet, and Z plus jet
(where the Z-boson decays to $\Pgmp\Pgmm$ or $\Pep\Pem$) events~\cite{Chatrchyan:2011ds}.
The jet energy resolution is
15\% at 10\GeV, 8\% at 100\GeV, and 4\% at 1\TeV~\cite{CMS-PAS-JME-10-003}.
Additional selection criteria are applied
to each event to remove spurious jet-like objects originating from isolated
noise patterns in certain HCAL regions.

Matching muons to tracks measured in the silicon tracker results in a relative
\pt resolution for muons with $20 <\pt < 100\GeV$ of 1.3--2.0\% in the barrel and
better than 6\% in the endcaps. The \pt resolution in the barrel is better than 10\% for
muons with \pt up to 1\TeV~\cite{cmsMuons}.
The mass resolution for $\cPZ\to\Pgm\Pgm$ decays is between 1.1\% and 1.9\% depending on
the pseudorapidity of each muon, for $\abs{\eta}<2.1$.
The mass resolution for $\cPZ \to \Pe \Pe$ decays when both electrons are in the
ECAL barrel (endcaps) is 1.6\% (2.6\%)~\cite{Chatrchyan:2013dga}.

 \section{\texorpdfstring{$\PH\to\Pgmp\Pgmm$}{H->mu+mu-} event selection}
\label{sec:evtSel}

Online collection of events is performed with a trigger that requires at least one
isolated muon candidate with \pt above 24\GeV in the pseudorapidity range $\abs{\eta} \le 2.1$.
In the offline selection, muon candidates are required to pass the
``Tight muon selection''~\cite{cmsMuons} and each muon trajectory is required to have an
impact parameter with respect to the primary vertex
smaller than 5\unit{mm} and 2\unit{mm} in the longitudinal and transverse directions, respectively.
They must also have $\pt>15\GeV$ and $\abs{\eta} \le 2.1$.

For each muon candidate, an isolation variable is constructed using the scalar
sum of the transverse-momentum of particles, reconstructed as PF
objects, within a cone centered on the muon.
The boundary of the cone is
$\Delta R=\sqrt{\smash[b]{(\Delta\eta)^2+(\Delta\phi)^2}}=0.4$ away from the muon, and
the \pt of the muon is not included in the sum.
While only charged particles associated with the primary vertex are taken into account,
a correction must be applied for contamination from neutral particles coming from pileup interactions.
On average, in inelastic proton-proton collisions, neutral pileup particles deposit half as much
energy as charged pileup particles.
The amount of energy coming from charged pileup particles is estimated as the
sum of the transverse momenta of charged tracks originating from vertices other than
the primary vertex, but still entering the isolation cone.
The neutral pileup energy in the isolation cone is then estimated to be 50\% of this value
and subtracted from the muon isolation variable.
A muon candidate is accepted
if the corrected isolation variable is less than 12\% of the muon \pt.

To pass the offline selection, events must contain a pair of opposite-sign muon candidates passing
the above selection, and the muon which triggered the event is required to have $\pt > 25$\GeV.
All combinations of opposite-sign pairs, where one of the muons triggers the event, are considered
as dimuon candidates in the dimuon invariant mass distribution analysis.  Each pair is effectively treated
as a separate event, and referred to as such for the remainder of this paper.
Less than 0.1\% of the SM Higgs boson events and
0.005\% of the background events in each category contain more than one pair of muons.

After selecting events with a pair of isolated opposite-sign muons, events
are categorized according to the properties of jets.
Jets reconstructed from PF objects are only considered if their \pt is greater than 30\GeV
and $\abs{\eta}<4.7$. A multivariate analysis (MVA) technique is used to discriminate between
jets originating from hard interactions and jets originating from pileup~\cite{JME-13-005}.

Dimuon events are classified into
two general categories: a 2-jet category and a 0,1-jet category.
The 2-jet category requires at least two jets, with $\pt>40$\GeV for
the leading jet and $\pt>30$\GeV for the subleading jet. A 2-jet event
must also have $\ptmiss<40$\GeV, where $\ptmiss$ is the
magnitude of the vector sum of the transverse momenta of the dimuon and dijet
systems.
The $\ptmiss$ requirement reduces the \ttbar contamination in the 2-jet category, since \ttbar decays also
include missing transverse momentum due to neutrinos.
All dimuon events not selected for the 2-jet category are placed into the
0,1-jet category where the signal is produced dominantly by GF.

The 2-jet category is further divided into VBF~Tight,
GF~Tight, and Loose subcategories.
The VBF Tight category has a large S/B  ratio for VBF produced events.
It requires
$\Mjj>650$\GeV and $\abs{\Delta\etajj}>3.5$, where $\abs{\Delta\etajj}$
is the absolute value of the difference in pseudorapidity between the two leading jets.
For a SM Higgs boson with $\mh=125\GeV$, 79\% of the signal events in this category
are from VBF production.
Signal events in the 2-jet category that do not pass the VBF~Tight criteria mainly arise from
GF events, which contain two jets from initial-state radiation.
The GF~Tight category captures these events by requiring
the dimuon transverse momentum (\ptmumu) to be greater than 50\GeV
and $\Mjj>250$\GeV.
To further increase the sensitivity of this search, 2-jet events that fail
the VBF Tight and GF Tight criteria are still retained in a third
subcategory called 2-jet Loose.

In the 0,1-jet category,
events are split into two subcategories based on the value of
\ptmumu. The most sensitive subcategory is 0,1-jet Tight which requires \ptmumu
greater than 10\GeV, while the events with \ptmumu less than 10\GeV are
placed in the 0,1-jet Loose subcategory. The S/B ratio
is further improved by categorizing events based on the dimuon invariant mass
resolution as follows. Given the narrow Higgs boson decay width, the mass
resolution fully determines the shape of the signal peak.
The dimuon mass resolution is dominated by the muon
\pt resolution, which worsens with increasing $\abs{\eta}$~\cite{cmsMuons}.
Hence, events are further sorted into subcategories
based on the $\abs{\eta}$ of each muon and are labeled as ``barrel'' muons (B) for
$\abs{\eta}<0.8$, ``overlap'' muons (O) for $0.8\leq\abs{\eta}<1.6$, and ``endcap'' muons
(E) for $1.6\leq\abs{\eta}<2.1$. The 0,1-jet dimuon events are then assigned, within
the corresponding Tight and Loose categories, to all possible dimuon $\abs{\eta}$
combinations.  The dimuon mass resolution for each category is shown in Table~\ref{tab:nEvts}.
  Due to the limited size of the data samples, the 2-jet
subcategories are not split into further subcategories according to the
muon \pt resolution.  This leads to a total of fifteen subcategories:
three 2-jet subcategories, six 0,1-jet Tight subcategories,
and six 0,1-jet Loose subcategories.

\begin{table*}[!hbtp]
  \centering
    \topcaption{ \label{tab:nEvts}
        Details regarding each of the \hmm categories.
        The top half of the table refers to the $5.0\pm0.1$\fbinv at 7\TeV, while the bottom
        half refers to the $19.7\pm0.5$\fbinv at 8\TeV.
        Each row lists the category name,
        FWHM of the signal peak,
        acceptance times selection efficiency ($A\epsilon$) for GF,
        $A\epsilon$ for VBF, $A\epsilon$ for VH,
        expected number of SM signal events in the category for
        $\mh=125\GeV$ ($N_\mathrm{S}$),
        number of background events within a FWHM-wide window centered on 125\GeV
        estimated by a signal plus background fit to the data ($N_\mathrm{B}$),
        number of observed events within a FWHM-wide window centered
        on 125\GeV ($N_\mathrm{Data}$), systematic uncertainty
        to account for the parameterization of the background ($N_\mathrm{P}$), and
        $N_\mathrm{P}$ divided by the statistical uncertainty on the fitted
        number of signal events ($N_\text{P}/\sigma_\text{Stat}$).
        The expected number of SM signal events is
        $N_\mathrm{S} = \mathcal{L}\times (\sigma  \BF  A  \epsilon)_\mathrm{GF}+\mathcal{L}\times (\sigma  \BF  A \epsilon)_\mathrm{VBF}+\mathcal{L}\times (\sigma  \BF  A  \epsilon)_\mathrm{VH}$,
        where $\mathcal{L}$ is the integrated luminosity and $\sigma \BF$ is the SM cross section
        times branching fraction.
     }
     \begin{tabular}{ld{1.1}d{1.1}d{1.1}d{1.1}d{2.2}d{5.1}d{5.1}d{2.2}d{2.0}} \hline
      & \mc{FWHM}
     & \multicolumn{3}{c}{$A\epsilon$ [\%]} & & & &
     & \mc{$N_\text{P}/\sigma_\text{Stat}$}   \\ \cline{3-5}
     Category & \multicolumn{1}{c}{[\GeVns{}]}
        & \mc{GF} & \mc{VBF} & \mc{VH}
        & \mc{$N_\mathrm{S}$}
        & \mc{$N_\mathrm{B}$}
        & \mc{$N_\mathrm{Data}$}
        & \mc{$N_\mathrm{P}$}
        & \mc{[\%]} \\ \hline
     0,1-jet Tight BB & 3.4 &   9.7 &   8.1 &   8.9 & 1.83 & 226.4 &  245 & 22.5 &  101 \\
     0,1-jet Tight BO & 4.0 &  14.0 &  11.0 &  13.0 & 2.56 & 470.3 &  459 & 42.4 &  121 \\
     0,1-jet Tight BE & 4.4 &   4.9 &   3.8 &   4.8 & 0.92 & 234.8 &  235 & 16.6 &   65 \\
     0,1-jet Tight OO & 4.8 &   5.2 &   3.9 &   4.9 & 0.97 & 226.5 &  236 & 11.5 &   52 \\
     0,1-jet Tight OE & 5.3 &   4.0 &   3.0 &   4.2 & 0.75 & 237.5 &  228 & 26.5 &  106 \\
     0,1-jet Tight EE & 5.9 &   0.9 &   0.7 &   1.0 & 0.17 &  71.4 &   57 & 11.4 &   97 \\
     0,1-jet Loose BB & 3.2 &   2.2 &   0.1 &   0.1 & 0.38 & 151.4 &  127 & 17.2 &   95 \\
     0,1-jet Loose BO & 3.9 &   3.1 &   0.2 &   0.2 & 0.52 & 307.0 &  291 & 18.9 &   71 \\
     0,1-jet Loose BE & 4.2 &   1.2 &   0.1 &   0.1 & 0.20 & 148.7 &  178 & 19.1 &  102 \\
     0,1-jet Loose OO & 4.5 &   1.2 &   0.1 &   0.1 & 0.20 & 144.7 &  143 & 19.1 &  113 \\
     0,1-jet Loose OE & 5.1 &   1.0 &   0.1 &   0.1 & 0.16 & 160.1 &  159 & 16.1 &   75 \\
     0,1-jet Loose EE & 5.8 &   0.2 &   0.0 &   0.0 & 0.03 &  41.6 &   39 &  5.6 &   51 \\
     2-jet VBF Tight  & 4.4 &   0.1 &   8.7 &   0.0 & 0.14 &   1.3 &    2 &  0.5 &   24 \\
     2-jet GF Tight   & 4.5 &   0.5 &   7.9 &   0.5 & 0.20 &  12.9 &   16 &  1.7 &   27 \\
     2-jet Loose      & 4.3 &   2.1 &   6.2 &  10.2 & 0.53 &  66.2 &   78 &  8.4 &   64 \\
     Sum of categories&\mcd &  50.3 &  53.9 &  48.1 & 9.56 & 2500.8& 2493 & \mcd & \mcd \\
     \hline
     0,1-jet Tight BB & 3.9 &   9.6 &   7.1 &   8.5 & 8.87 & 1208.0 & 1311 &  40.8 &   73 \\
     0,1-jet Tight BO & 4.4 &  13.0 &  10.0 &  13.0 &12.45 & 2425.3 & 2474 & 102.2 &  127 \\
     0,1-jet Tight BE & 4.7 &   4.9 &   3.4 &   4.6 & 4.53 & 1204.8 & 1212 &  63.8 &  111 \\
     0,1-jet Tight OO & 5.0 &   5.3 &   3.6 &   5.0 & 4.90 & 1112.7 & 1108 &  39.0 &   71 \\
     0,1-jet Tight OE & 5.5 &   4.1 &   2.8 &   4.2 & 3.85 & 1162.1 & 1201 & 151.1 &  251 \\
     0,1-jet Tight EE & 6.4 &   0.9 &   0.6 &   1.1 & 0.85 &  350.8 &  323 &  34.2 &  107 \\
     0,1-jet Loose BB & 3.7 &   2.1 &   0.1 &   0.1 & 1.73 &  715.4 &  697 &  40.2 &   94 \\
     0,1-jet Loose BO & 4.3 &   2.9 &   0.2 &   0.2 & 2.41 & 1436.4 & 1432 &  85.5 &  158 \\
     0,1-jet Loose BE & 4.5 &   1.1 &   0.1 &   0.1 & 0.90 &  725.9 &  782 &  74.9 &  166 \\
     0,1-jet Loose OO & 4.9 &   1.1 &   0.1 &   0.1 & 0.96 &  727.4 &  686 &  33.2 &   74 \\
     0,1-jet Loose OE & 5.5 &   0.9 &   0.1 &   0.1 & 0.76 &  791.8 &  832 &  78.2 &  158 \\
     0,1-jet Loose EE & 6.2 &   0.2 &   0.0 &   0.0 & 0.18 &  218.5 &  209 &  18.9 &   87 \\
     2-jet VBF Tight  & 5.0 &   0.2 &  11.0 &   0.0 & 0.95 &   10.6 &    8 &   1.6 &   35 \\
     2-jet GF Tight   & 5.1 &   0.7 &   8.4 &   0.6 & 1.14 &   74.8 &   76 &  11.8 &   88 \\
     2-jet Loose      & 4.7 &   2.4 &   6.3 &  10.4 & 2.90 &  431.7 &  387 &  25.3 &   73 \\
     Sum of categories&\mcd &  49.4 &  53.8 &  48.0 &47.38 &12596.2 &12738 &  \mcd & \mcd \\
     \hline
    \end{tabular}

\end{table*}

 \section{\texorpdfstring{$\mathrm{H}\to\mu^+\mu^-$}{H->mu+mu-} event selection efficiency}
\label{sec:selEff}

While the background shape and normalization are obtained from
data, the selection efficiency for signal events has to be determined using MC
simulation.
For the GF and VBF production modes, signal samples are produced
using the \textsc{powheg--box} next-to-leading-order (NLO) generator~\cite{powheg,powhegGF,powhegVBF}
interfaced with \PYTHIA~6.4.26~\cite{pythia} for parton showering.
VH samples are produced using \HERWIGpp~\cite{herwigpp}
and its integrated implementation of the NLO POWHEG method.

These samples are then passed through a simulation of the CMS detector,
based on \GEANTfour~\cite{geant4}, that has been extensively validated on
both 7 and 8\TeV data.
This validation includes a comparison of data with MC simulations of the
Drell--Yan plus jets and \ttbar plus jets backgrounds produced using \MADGRAPH~\cite{madgraph} interfaced with
\PYTHIA~6.4.26 for parton showering.
In all categories, the simulated $\Mmumu$ spectra
agree well with the data, for $110 < \Mmumu < 160\GeV$.
Scale factors related to muon identification, isolation,
and trigger efficiency are applied to each simulated signal sample to correct for
discrepancies between the detector simulation and data.  These scale factors are
estimated using the ``tag-and-probe'' technique~\cite{cmsMuons}.  The detector
simulation and data typically agree to within 1\% on the muon identification efficiency,
to within 2\% on the muon isolation efficiency, and to within 5\% on the muon trigger efficiency.

The overall acceptance times selection efficiency for the \hmm
signal depends on the mass of the Higgs boson. For a Higgs boson mass of
125\GeV, the acceptance times selection efficiencies are shown in
Table~\ref{tab:nEvts}.

 \section{\texorpdfstring{$\mathrm{H}\to\mu^+\mu^-$}{H->mu+mu-} systematic uncertainties}
\label{sec:systUnc}

Since the statistical analysis is performed on the dimuon invariant mass spectrum, it is necessary
to categorize the sources of systematic uncertainties into
``shape'' uncertainties that change the shape of the dimuon invariant mass
distribution, and ``rate'' uncertainties that affect the overall signal yield
in each category.

The only relevant shape uncertainties for the signal are related to the knowledge of the muon momentum scale
and resolution and they affect the width of the signal peak by 3\%.
The signal shape is parameterized by a double-Gaussian (see Section~\ref{sec:results})
and this uncertainty is applied by constraining the width of the narrower Gaussian.
The probability density function used to constrain this nuisance parameter in the limit setting procedure
is itself a Gaussian with its mean set to the nominal value and its width set
to 3\% of the nominal value.

Rate uncertainties in the signal yield are evaluated separately for each Higgs boson
production process and each centre-of-mass energy.
These uncertainties are applied using log-normal probability density functions as described
in Ref.~\cite{HiggsStats}.
Table~\ref{tab:hmmSyst} shows the relative systematic uncertainties in the
signal yield for $\mh{}=125\GeV$, with more detail given below.

\begin{table*}[!hbtp]
  \centering
    \topcaption{ \label{tab:hmmSyst}
    The relative systematic uncertainty in the \hmm{} signal yield is listed for
    each uncertainty source.  Uncertainties are shown for the GF and VBF Higgs
    boson production modes.  The systematic uncertainties vary depending on the
    category and centre-of-mass energy.
     }
\begin{tabular}{ l c c } \hline
Source                                           & GF [\%]   & VBF [\%]   \\ \hline
Higher-order corrections~\cite{LHCHXSWG3}        & 1--25      & 1--7        \\
PDF~\cite{LHCHXSWG3}                             & 11        & 5          \\
PS/UE                                            & 6--60      & 2--15       \\
\BF(\hmm{})~\cite{LHCHXSWG3}                     & 6         & 6          \\
Integrated luminosity~\cite{lumi_2011,lumi_2013} & 2.2--2.6   & 2.2--2.6    \\
MC statistics                                    & 1--8       & 1--8        \\
Muon efficiency                                  & 1.6       & 1.6        \\
Pileup                                           & $<1$--5    & $<1$--2     \\
Jet energy resolution                            & 1--3       & 1--2        \\
Jet energy scale                                 & 1--8       & 2--6        \\
Pileup jet rejection                             & 1--4       & 1--4        \\ \hline
\end{tabular}
\end{table*}

To estimate the theoretical uncertainty in the signal
production processes due to neglected higher-order
quantum corrections, the renormalization and
factorization scales are varied simultaneously by a
factor of two up and down from their nominal values.
This leads to an uncertainty in
the cross section and acceptance times efficiency which depends
on the mass of the Higgs boson.
The uncertainty is largest in the 2-jet VBF Tight and GF Tight categories, and
smallest in the 0,1-jet Tight categories.

Uncertainty in the knowledge of the parton distribution functions (PDFs) also
leads to uncertainty in the signal production process.
This uncertainty is estimated using the PDF4LHC
prescription~\cite{Alekhin:2011sk,Botje:2011sn}
and the CT10~\cite{Lai:2010vv},
MSTW2008~\cite{Martin:2009iq}, and NNPDF 2.3~\cite{Ball:2010de}
PDF sets provided by the \textsc{lhapdf} package version
5.8.9~\cite{LHAPDF}.
The value of the uncertainty depends on the mass of the Higgs boson, while the
dependence on the category is small.

Uncertainty in the modeling of the parton showers and underlying event
activity (PS/UE) may affect the kinematics of selected jets. This uncertainty is estimated
by comparing various tunes of the relevant \PYTHIA parameters. The
D6T~\cite{tune_z2}, P0~\cite{tune_P0}, ProPT0, and ProQ20~\cite{tune_proq20}
tunes are compared with the Z2*~\cite{tune_z2} tune, which is the nominal
choice.
The uncertainty is larger in the 2-jet categories than in the 0,1-jet
categories.  Large uncertainties in the 2-jet categories are expected for the
GF production mode, since two-jet events are simulated solely by parton
showering in the \POWHEG--\PYTHIA NLO samples.

Misidentification of ``hard jets'' (jets originating from the hard interaction) as ``pileup jets'' (jets originating
from pileup interactions) can lead to migration of signal events from the 2-jet
category to the 0,1-jet category.
Events containing a Z-boson, tagged by its dilepton decay, recoiling against a jet
provide a pure source of hard jets similar to the Higgs boson signal.
Data events may then be used to estimate the misidentification
rate of the MVA technique used to discriminate between hard jets and pileup jets using data~\cite{JME-13-005}.
A pure source of hard jets is found by selecting events with $\pt^{\cPZ}>30\GeV$
and jets where $\abs{\Delta \phi(Z,\mathrm{j})} > 2.5$ and
$0.5<\pt^{\mathrm{j}}/\pt^{\cPZ}<1.5$.
The misidentification rate of these jets as pileup jets is compared in data and simulation,
and the difference taken as a systematic uncertainty.

There are several additional uncertainties.  The theoretical uncertainty in the
branching fraction to $\Pgmp\Pgmm$ is taken from Ref.~\cite{LHCHXSWG3}, and
depends on the Higgs boson mass.  The uncertainty in the luminosity is directly
applied to the signal yield in all categories.  The signal yield uncertainty
due to the limited size of the simulated event samples depends on the category,
and is listed as ``MC statistics'' in Table~\ref{tab:hmmSyst}.  There is a
small uncertainty associated with the ``tag-and-probe'' technique used to
determine the data to simulation muon efficiency scale factors~\cite{cmsMuons}.
This uncertainty is labeled ``Muon efficiency'' in Table~\ref{tab:hmmSyst}.  A
systematic uncertainty in the knowledge of the pileup multiplicity is evaluated
by varying the total cross section for inelastic proton-proton collisions.  The
acceptance and selection efficiency of the jet-based selections are affected by
uncertainty in the jet energy resolution and absolute jet energy scale
calibration~\cite{Chatrchyan:2011ds}.

For VH production, only rate uncertainties in the production cross section
due to quantum corrections and PDFs are considered.  They are 3\% or less~\cite{LHCHXSWG3}.

When estimating each of the signal yield uncertainties, attention is paid to the sign of the yield
variation in each category.  Categories that vary in the same direction are considered fully correlated
while categories that vary in opposite directions are considered anticorrelated.
These correlations are considered between all categories at both beam energies for all of the signal
yield uncertainties except for the luminosity uncertainty and the uncertainty caused by the
limited size of the simulated event samples.
The luminosity uncertainty is considered fully
correlated between all categories, but uncorrelated between the two centre-of-mass energies.
The MC simulation statistical uncertainty is considered uncorrelated between all categories
and both centre-of-mass energies.

To account for the possibility that the nominal background parameterization may imperfectly
describe the true background shape, an additional systematic uncertainty is included.
This uncertainty is implemented as a floating additive contribution to the number
of signal events, constrained by a Gaussian probability density function with mean set to zero and
width set to the systematic uncertainty.
This systematic uncertainty is estimated by checking the
bias in terms of the number of signal events that are found when fitting the signal plus
nominal background model (see Section~\ref{sec:results}) to pseudo-data generated
from various alternative background models, including polynomials, that were fit to data.
Bias estimates are performed for Higgs boson mass points from 120 to 150\GeV.
The uncertainty estimate is then taken as the maximum absolute value of the bias of all of the
mass points and all of the alternative background models.  It is then applied uniformly
to all Higgs boson masses.
The estimates of the uncertainty in the parameterization of the background ($N_\text{P}$)
are shown in Table~\ref{tab:nEvts} for each category.
The effect of this systematic uncertainty is larger than all of the others. The expected
limit (see Section~\ref{sec:results}) would be 20\% lower at $\mh = 125\GeV$ without the
systematic uncertainty in the parameterization of the background.

 \section{\texorpdfstring{$\mathrm{H}\to\mu^+\mu^-$}{H->mu+mu-} results}
\label{sec:results}

\begin{figure}[!hbtp]
  \centering
    \includegraphics[width=0.49\textwidth]{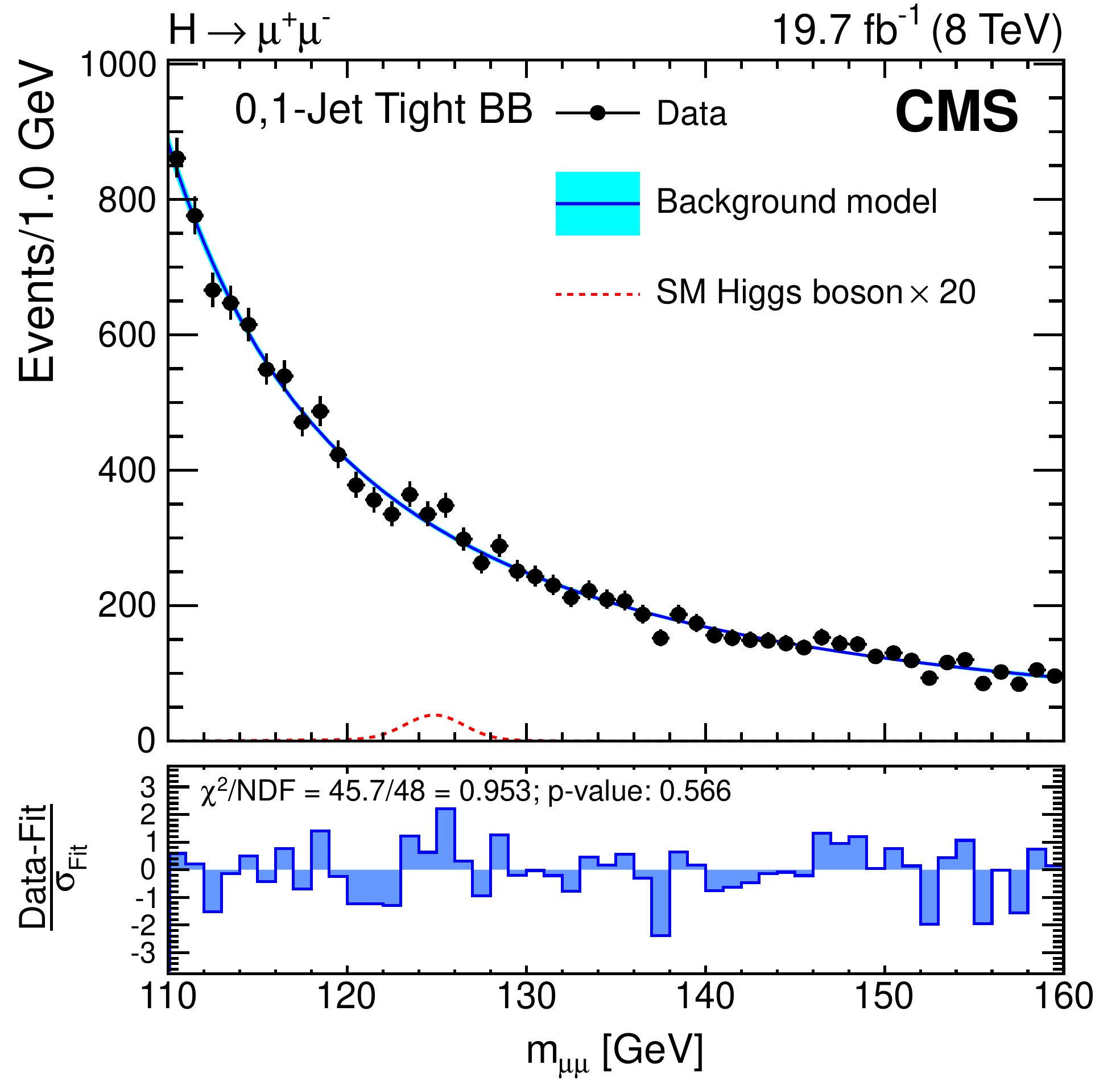}
    \includegraphics[width=0.49\textwidth]{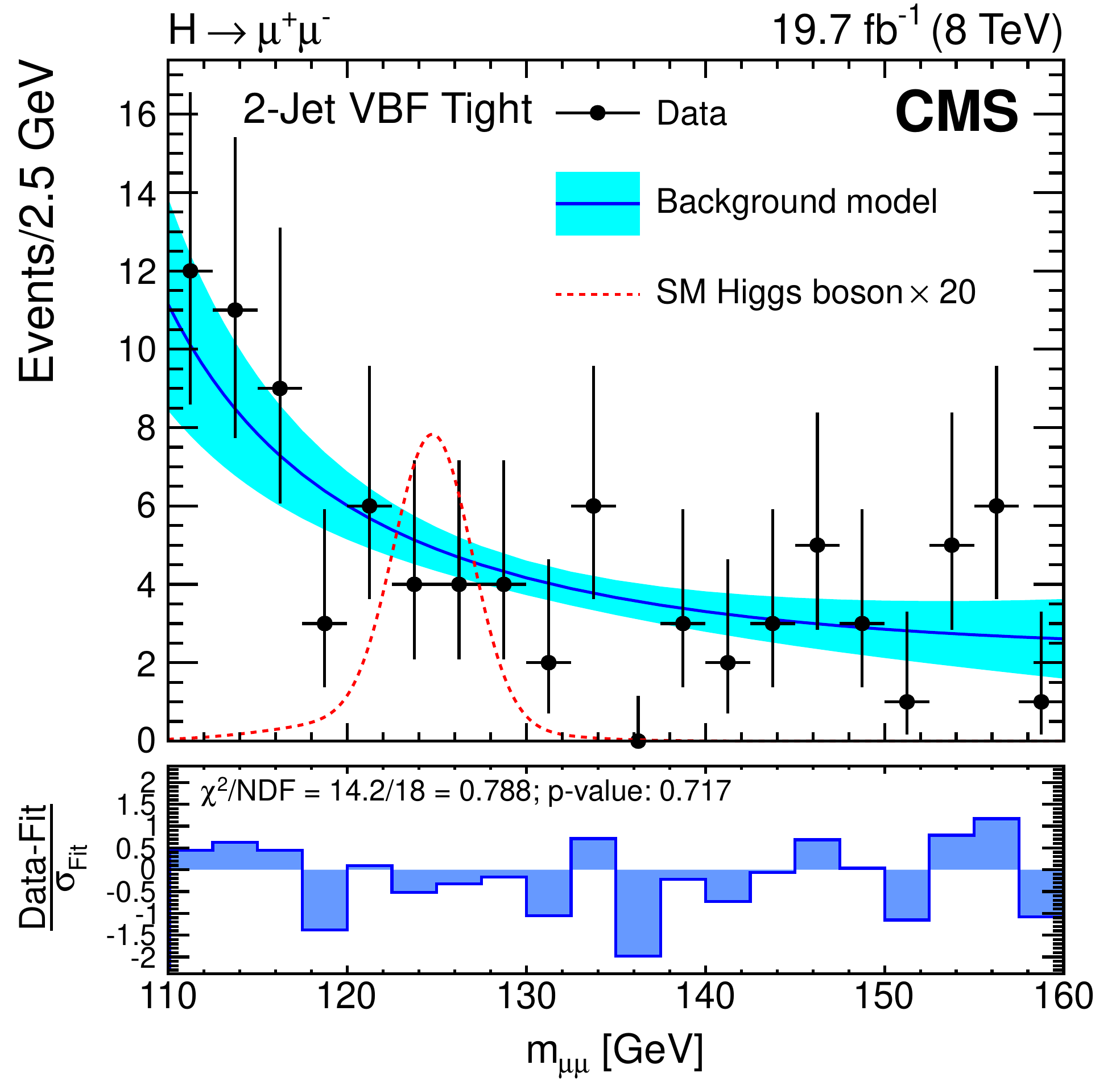}
    \caption{ The dimuon invariant mass at 8\TeV and the background model are shown
      for the 0,1-jet Tight category when both muons are reconstructed
      in the barrel (\cmsLeft) and the 2-jet VBF Tight category (\cmsRight).
      A best fit of the background model (see text) is shown
      by a solid line, while its fit uncertainty is represented by a lighter band.
      The dotted line illustrates the expected SM Higgs boson signal enhanced by a factor of 20,
      for $\mh=125\GeV$.
      The lower histograms show the residual for each bin (Data-Fit) normalized by
      the Poisson statistical uncertainty of the background model ($\sigma_\mathrm{Fit}$).
      Also given are the sum of squares
      of the normalized residuals ($\chi^2$) divided by the number of degrees of freedom (NDF) and
     the corresponding $p$-value assuming the sum follows the $\chi^2$ distribution.}
    \label{fig:bakShapeData8TeV_pas}

\end{figure}

To estimate the signal rate, the dimuon invariant mass (\Mmumu)
spectrum is fit with the sum of parameterized signal and background shapes.
This fit is performed simultaneously in all of the categories.
Since  in the mass range of interest the natural width of the Higgs boson is narrower than the
detector resolution, the \Mmumu shape
is only dependent on the detector resolution and QED
final state radiation. A double-Gaussian function is chosen to parameterize the shape of the
signal.
The parameters that specify the signal shape are estimated by fitting the double-Gaussian function
to simulated signal samples.  A separate set of signal shape parameters are used for each category.
The background shape, dominated by the Drell--Yan process, is modeled by a function, $f(\Mmumu)$, that is the sum
of a Breit--Wigner function and a 1/$\Mmumu^2$ term, to model the \cPZ-boson and photon contributions, both multiplied
by an exponential function to approximate the effect of the PDF on the \Mmumu distribution.
This function is shown in the following equation,
and involves the parameters $\lambda$, $\beta$, $\mz$, and $\Gamma$:
\ifthenelse{\boolean{cms@external}}{
\begin{multline}\label{eqn:bkg}
 f(\Mmumu) =  \beta C_1 \re^{-\lambda \Mmumu} \frac{1}{(\Mmumu-\mz)^2
 + \frac{\Gamma^2}{4}} \\+ (1-\beta) C_2 \re^{-\lambda \Mmumu} \frac{1}{\Mmumu^2}.
\end{multline}
}
{
\begin{equation}\label{eqn:bkg}
 f(\Mmumu) =  \beta C_1 \re^{-\lambda \Mmumu} \frac{1}{(\Mmumu-\mz)^2 + \frac{\Gamma^2}{4}} + (1-\beta) C_2 \re^{-\lambda \Mmumu} \frac{1}{\Mmumu^2}.
\end{equation}
}
The coefficients $C_1$ and $C_2$ are set to ensure
the integral of each of the two terms is normalized to unity in the \Mmumu fit range, 110 to 160\GeV.
Each category uses a different set of background parameters.
Before results are extracted, the mass and width of the \cPZ-boson peak, $\mz$ and $\Gamma$, are estimated by
fitting a Breit--Wigner function to the \Z-boson mass peak region (88--94\GeV) in each category.
The other parameters, $\lambda$ and $\beta$, are fit simultaneously with
the amount of signal in the signal plus background fit.
Besides the Drell--Yan process, most of the remaining background events come from \ttbar production.  The
background parameterization has been shown
to fit the dimuon mass spectrum well, even when it includes a large \ttbar fraction.
Fits of the background model to data (assuming no signal contribution) are presented in
Fig.~\ref{fig:bakShapeData8TeV_pas} for the most sensitive categories: the
0,1-jet Tight category with both muons reconstructed in the barrel region and
the 2-jet VBF Tight category.

Results are presented in terms of the signal strength, which is the ratio
of the observed (or expected) $\sigma\BF$,
to that predicted in the SM for the \hmm process.  Results are also presented,
for $\mh=125\GeV$, in terms of $\sigma\BF$, and \BF.
No significant excess is observed. Upper limits at the 95\% CL
are presented using the $\mathrm{CL_s}$ criterion~\cite{CLS1,CLS2}.
They are calculated using an asymptotic profile likelihood ratio
method~\cite{cmsCombineTool,HiggsStats,AsymptoticLimits}
involving dimuon mass shapes for each signal process and for background.
 Systematic uncertainties are incorporated as nuisance parameters and treated
according to the frequentist paradigm~\cite{HiggsStats}.

\begin{figure}[!hbtp]
  \centering
   \includegraphics[width=0.49\textwidth]{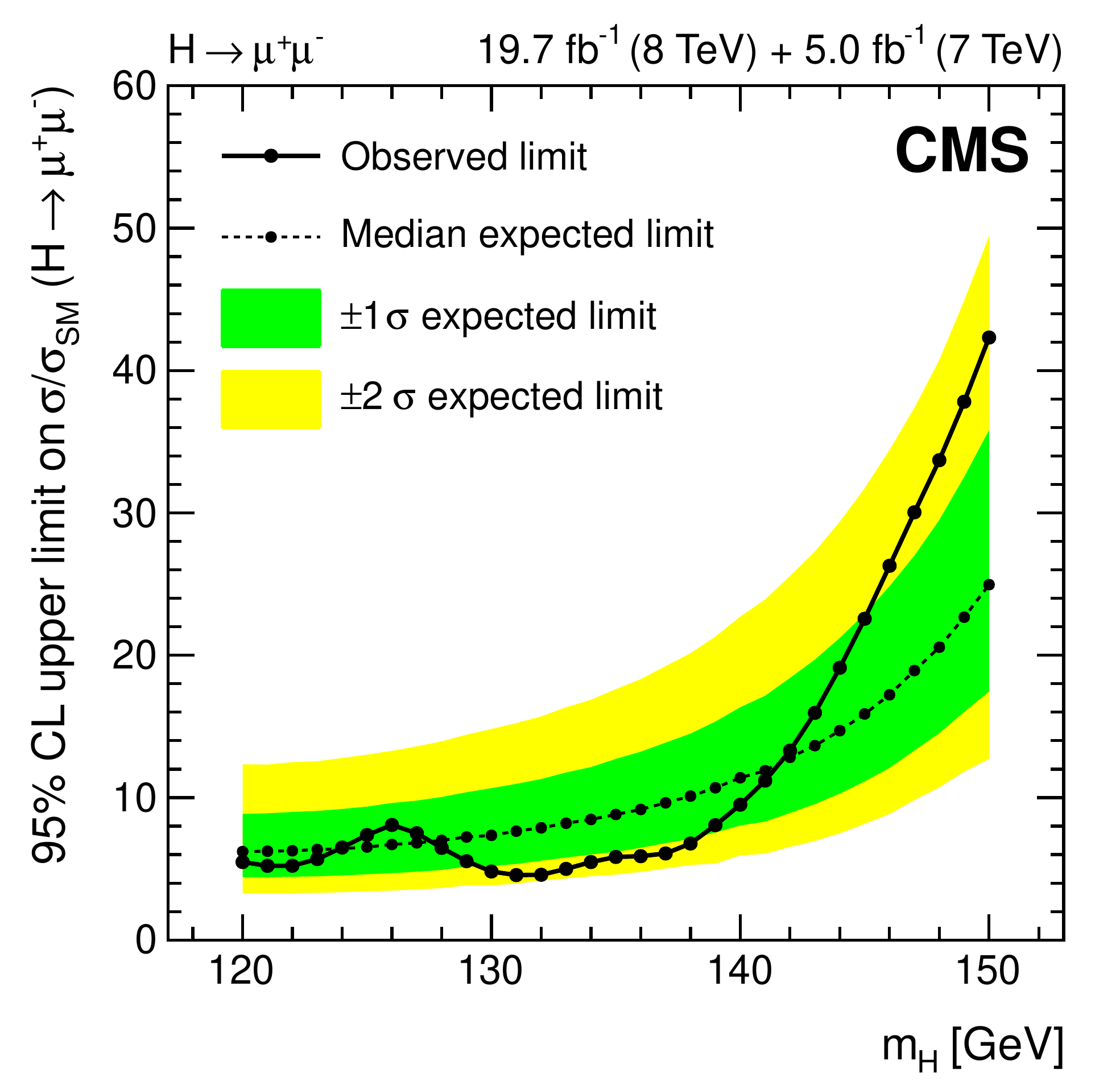}
     \caption{ Mass scan for the background-only expected and observed combined exclusion limits.}
     \label{fig:expectedLimitsMassScan_pas}

\end{figure}

Exclusion limits
 for Higgs boson masses from 120 to 150\GeV
are shown in Fig.~\ref{fig:expectedLimitsMassScan_pas}.
The observed 95\% CL upper limits on the signal strength at 125\GeV are
22.4 using the 7\TeV data and 7.0 using the 8\TeV data.
The corresponding background-only expected limits are
$16.6^{+7.3}_{-4.9}$ using the 7\TeV data and
$7.2^{+3.2}_{-2.1}$ using the 8\TeV data.
Accordingly, the combined observed limit for 7 and 8\TeV is 7.4,
 while the background-only expected limit is $6.5^{+2.8}_{-1.9}$.
This corresponds to an observed upper limit on $\BF(\hmm)$ of 0.0016,
assuming the SM cross section.
The best fit value of the signal strength for a Higgs boson mass of 125\GeV is
$0.8^{+3.5}_{-3.4}$.
We did not restrict the fit to positive values, to preserve the generality of
the result.

\begin{figure}[!hbtp]
\centering
    \includegraphics[width=0.49\textwidth]{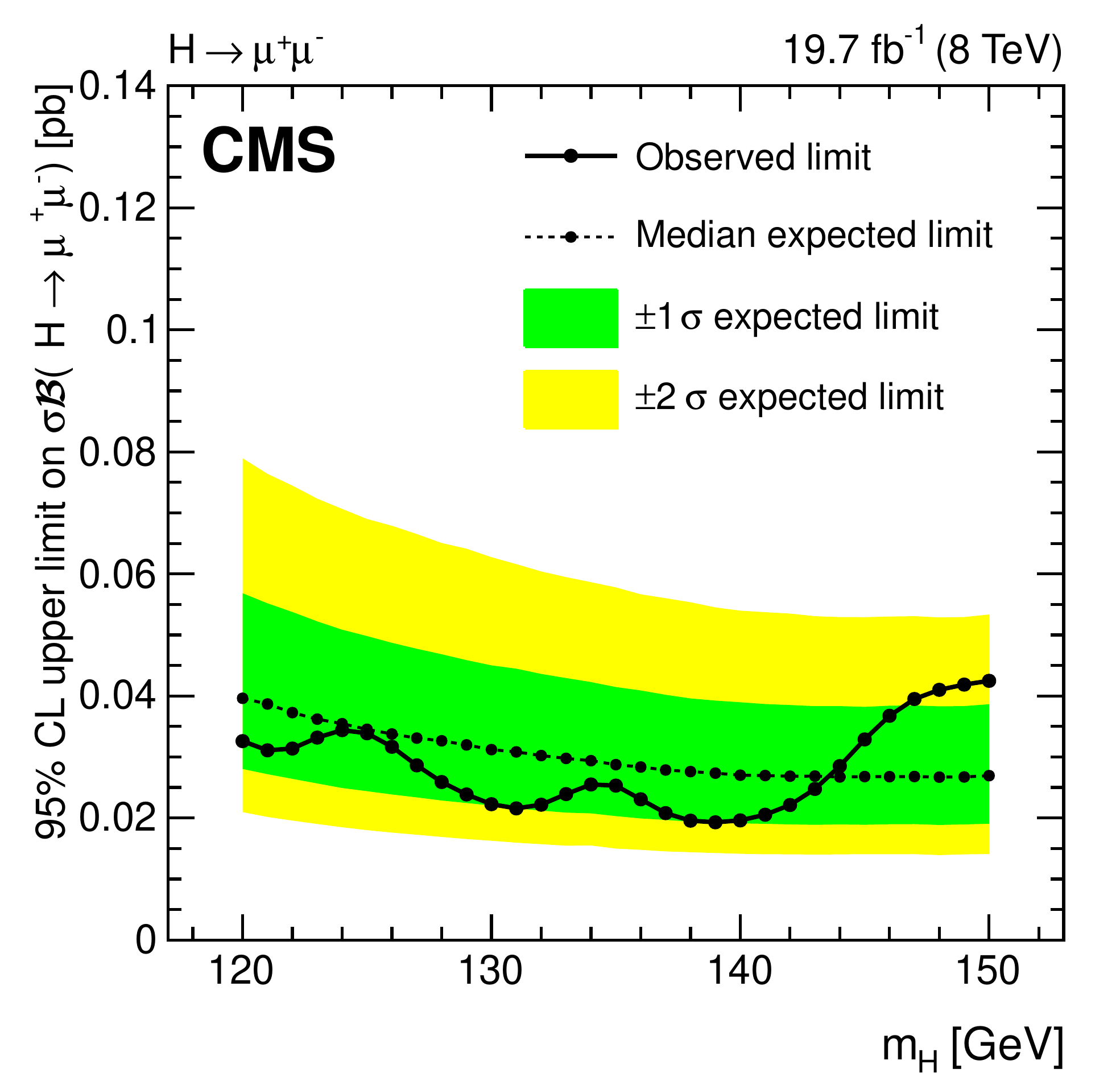}
    \includegraphics[width=0.49\textwidth]{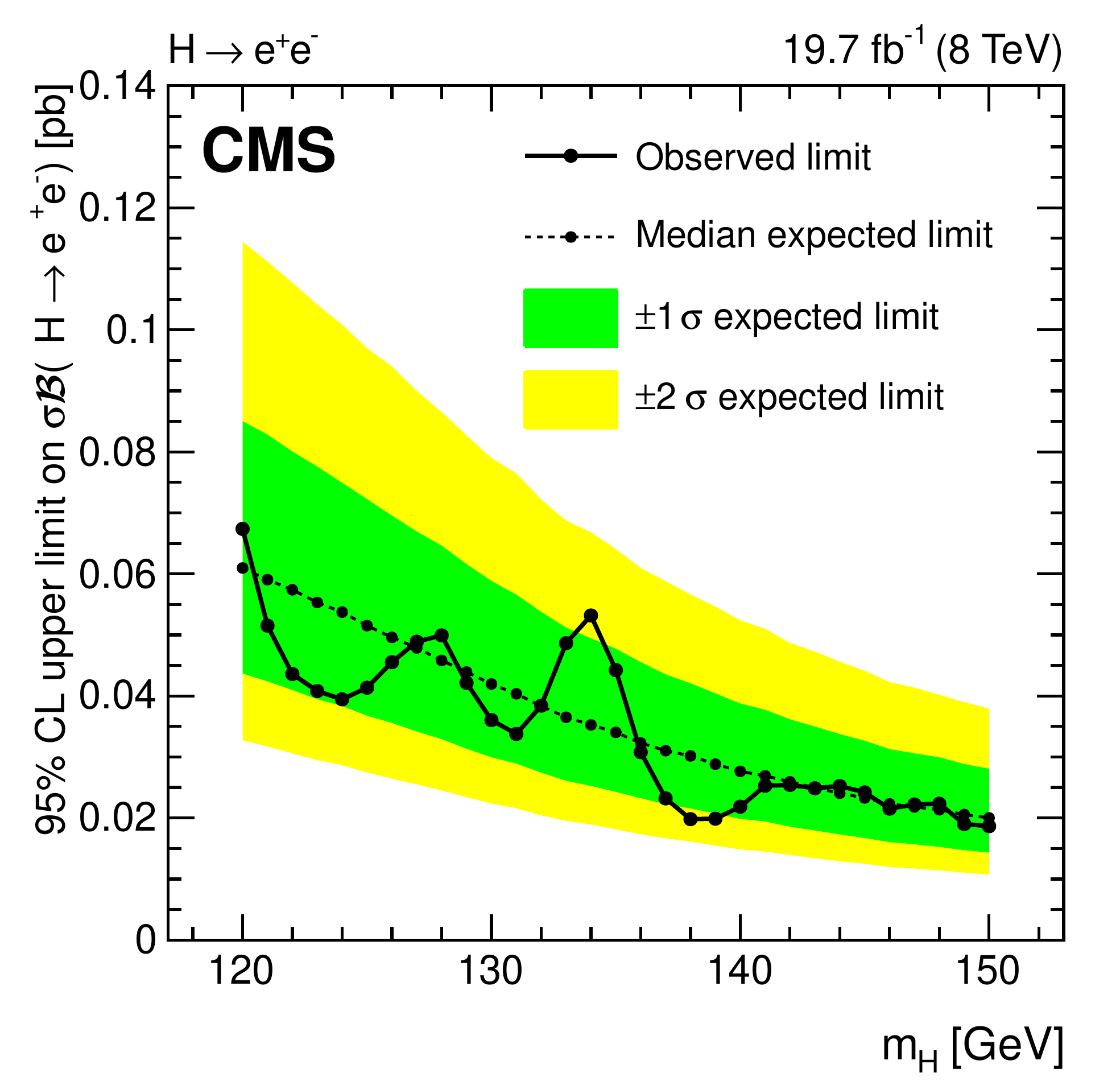}
    \caption{ \label{fig:xsbrLimits}
      Exclusion limits on $\sigma \BF$ are shown for
      \hmm (\cmsLeft),
      and for \hee (\cmsRight), both for 8\TeV.
      Theoretical uncertainties on the
      cross sections and branching fraction are omitted, and the relative contributions
      of GF, VBF, and VH are as predicted in the SM.
      }

\end{figure}

Exclusion limits in terms of $\sigma(\text{8\TeV})\BF$
using only 8\TeV data are shown in Fig.~\ref{fig:xsbrLimits} (\cmsLeft).
The relative contributions of GF, VBF, and VH are assumed to be as predicted
in the SM, and theoretical uncertainties on the cross sections and branching
fractions are omitted.
At 125\GeV,
the observed 95\% CL upper limit on $\sigma(\text{7\TeV})  \BF$ using only 7\TeV data
is 0.084\unit{pb}, while the background-only expected limit is 0.062$^{+0.026}_{-0.018}$\unit{pb}.
Using only 8\TeV data, the observed limit on $\sigma(\text{8\TeV})  \BF$
is  0.033\unit{pb}, while the background-only expected limit is 0.034$^{+0.014}_{-0.010}$\unit{pb}.

Exclusion limits on individual production modes may also be useful to constrain
BSM models that predict \hmm production dominated by a single mode.  Limits
are presented on the signal strength using a
combination of 7 and 8\TeV data and on $\sigma(8\TeV)  \BF$ using only the 8 TeV data.
The observed 95\% CL upper limit on the GF signal strength, assuming
the VBF and VH rates are zero, is 13.2, while the background-only expected
limit is $9.8^{+4.4}_{-2.9}$.    Similarly, the observed upper limit on the VBF
signal strength, assuming the GF and VH rates are zero, is 11.2, while the
background-only expected limit is $13.4^{+6.6}_{-4.2}$.
The observed upper limit on
$\sigma_\mathrm{GF}(8\TeV)  \BF$ is 0.056\unit{pb} and expected limit is
$0.045^{+0.019}_{-0.013}$\unit{pb}, using only 8\TeV data.
Similarly, the
observed upper limit on $\sigma_\mathrm{VBF}(8\TeV)  \BF$ is
0.0036\unit{pb} and the expected limit is $0.0050^{+0.0024}_{-0.0015}$\unit{pb}, using only 8\TeV data.

For $\mh=125\GeV$, an alternative \hmm analysis was performed to check the results of the main analysis.
It uses an alternative muon isolation variable based only on tracker information, an
alternative jet reconstruction algorithm (the jet-plus-track algorithm~\cite{JME-09-002}),
and an alternative event categorization.  The event categorization contains similar
2-jet categories to the main analysis, while separate categories are utilized for
0-jet and 1-jet events.  Dimuon mass resolution-based categories are not used,
but the 0-jet category does contain two subcategories separated by $\ptmumu$.
As in the main analysis, results are extracted by fitting signal and background shapes to the \Mmumu
spectra in each category, but unlike the main analysis,
$f(\Mmumu)=\exp(p_1\Mmumu)/(\Mmumu-p_2)^2$ is used as the background shape.
The systematic uncertainty on the parameterization of the background is estimated and applied
in the same way as in the main analysis.
For the alternative analysis, the observed (expected) 95\% CL upper limit on the signal strength
is 7.8 (6.5$^{+2.8}_{-1.9}$) for the combination of 7\TeV and 8\TeV data and $\mh=125\GeV$.
The observed limits of both the main and alternative analyses are within
one standard deviation of their respective background-only expected limits, for $\mh=125\GeV$.

\section{Search for Higgs boson decays to \texorpdfstring{$\mathrm{e^+e^-}$}{e+e-}}
\label{sec:hee}

In the SM, the branching fraction of the Higgs boson into \ee is tiny,
because the fermionic decay width is proportional to the mass of the fermion squared.
This leads to poor sensitivity to SM production for this search when compared to the search for \hmm.
On the other hand, the sensitivity in terms of $\sigma\BF$ is similar to \hmm, because
dielectrons and dimuons share similar
invariant mass resolutions, selection efficiencies, and
backgrounds.
Since the sensitivity to the SM rate of \hee is so poor,
an observation of the newly discovered particle decaying to \ee with the current integrated luminosity
would be evidence of physics beyond the standard model.

In a similar way to the \hmm analysis, a search in the \Mee spectrum is performed for a narrow peak
over a smoothly falling background.  The irreducible background is dominated by Drell--Yan production,
with smaller contributions from $\ttbar$ and diboson production.
Misidentified electrons make up a reducible background that is highly suppressed by
the electron identification criteria. The reducible \hgamgam background is estimated from simulation to
be negligible compared to other backgrounds, although large compared to the SM \hee signal.
The overall background shape and normalization
are estimated by fitting the observed \Mee spectrum in data, assuming a smooth functional form,
while the signal acceptance times selection efficiency is estimated from simulation.
The analysis is performed only on proton-proton collision data collected at 8\TeV,
corresponding to an integrated luminosity of $19.7\pm0.5$\fbinv.

The trigger selection requires two electrons,
one with transverse energy, \et, greater
than 17\GeV and the other with \et greater than 8\GeV.
These electrons are required to be isolated with respect to additional energy
deposits in the ECAL, and to pass selections on the ECAL cluster shape.
In the offline selection, electrons are required to be inside the
ECAL fiducial region: $\abs{\eta} < 1.44$ (barrel)
or $1.57 < \abs{\eta} < 2.5$ (endcaps).
Their energy is estimated by the same multivariate regression technique used in the
CMS \hzz analysis~\cite{HIG-13-002}, and their \et is required to be greater than 25\GeV.
Electrons are also required to satisfy standard CMS identification and isolation requirements,
which correspond to a single electron efficiency of around 90\% in the barrel
and 80\% in the endcaps~\cite{CMS-DP-2013-003}.

To improve the sensitivity of the search we separate the sample into four
distinct categories: two 0,1-jet categories and two for which a pair of jets is required.
The two 2-jet categories are designed to select events produced via the VBF process.
The two jets are required to have an invariant mass greater than 500\,(250)\GeV for the
2-jet Tight (Loose) category, $\pt > 30\,(20)\GeV$,
$\abs{\Delta\etajj} > 3.0$,
$\abs{\Delta\phi(\mathrm{jj},\Pep\Pem)}>2.6$, and
$\abs{z}=\abs{\eta(\Pep\Pem)-[\eta(\mathrm{j_1})+\eta(\mathrm{j_2})]/2}< 2.5$~\cite{zeppenfeld}.
The cut on $z$  ensures that the dielectron is produced centrally in the dijet reference frame,
which helps to enhance the VBF signal over the Drell--Yan background.
More details on the selection can be found in Ref.~\cite{HIG-13-001}.
The rest of the events are classified into two 0,1-jet categories. To exploit the better energy
resolution of electrons in the barrel region, these categories are defined as:
both electrons in the ECAL barrel (0,1-jet BB)
or at least one of them in the endcap (0,1-jet Not BB).  For each category, the FWHM of the expected signal peak,
expected number of SM signal events for $\mh=125\GeV$, acceptance times selection efficiency,
number of background events near 125\GeV, and number of data events near 125\GeV are shown
in Table~\ref{tab:nEvtsHee}.

Data have been compared to the simulated Drell--Yan and \ttbar background samples
described in Section~\ref{sec:selEff}. In all categories, the dielectron invariant mass spectra
from 110 to 160\GeV agree well, and the normalizations agree within 4.5\%.
Using simulation, the reducible background of \hgamgam events has also been estimated.
For $\mh=125$\GeV, 0.23 SM \hgamgam events are expected to pass the dielectron selection
compared to about $10^{-3}$ events for the SM \hee signal.
While this background is much larger than the SM \hee signal, it is negligible compared
to the Drell--Yan and \ttbar backgrounds in each category.

\begin{figure}[!hbtp]
 \centering
 \includegraphics[width=0.49\textwidth]{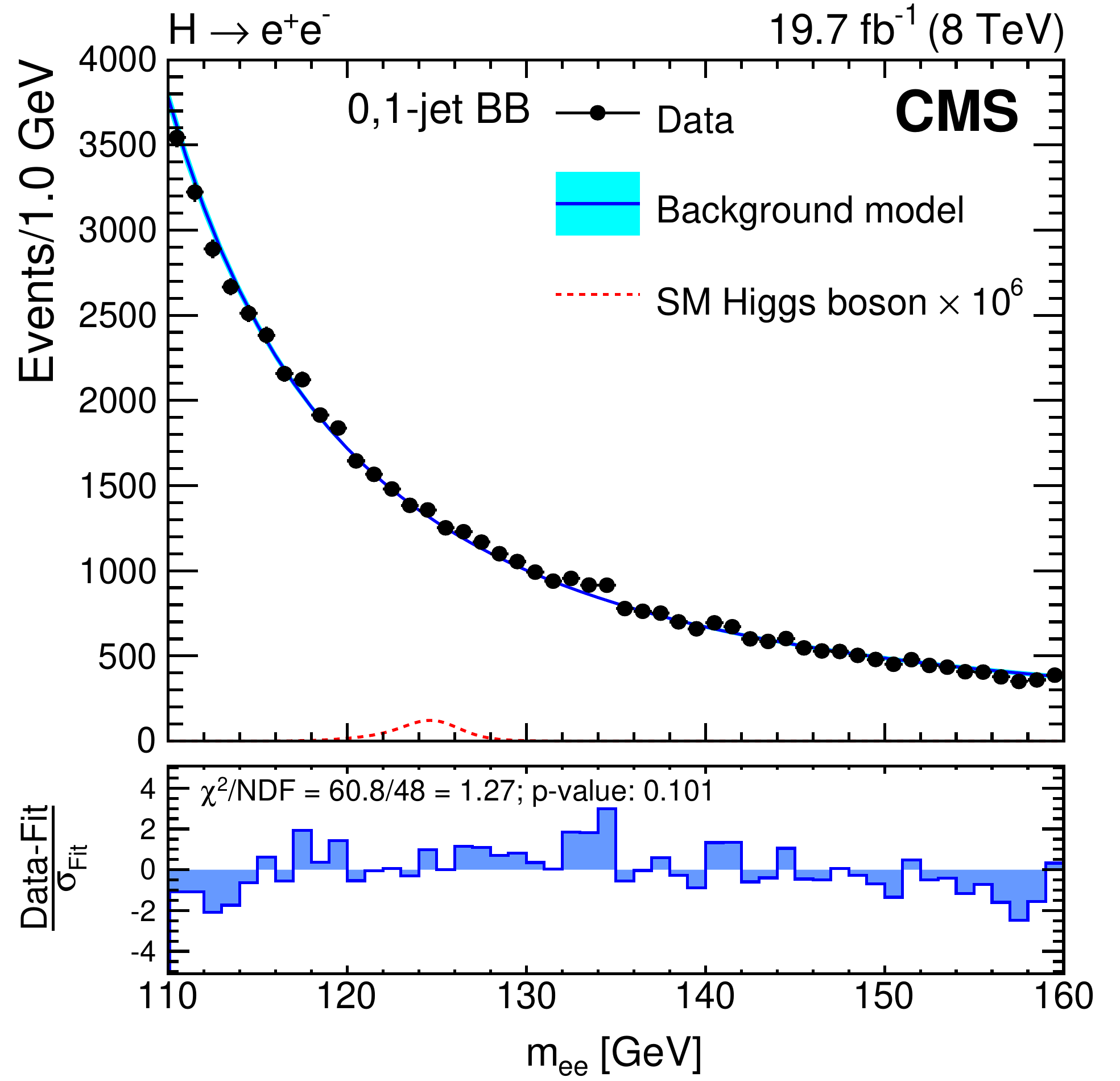}
 \includegraphics[width=0.49\textwidth]{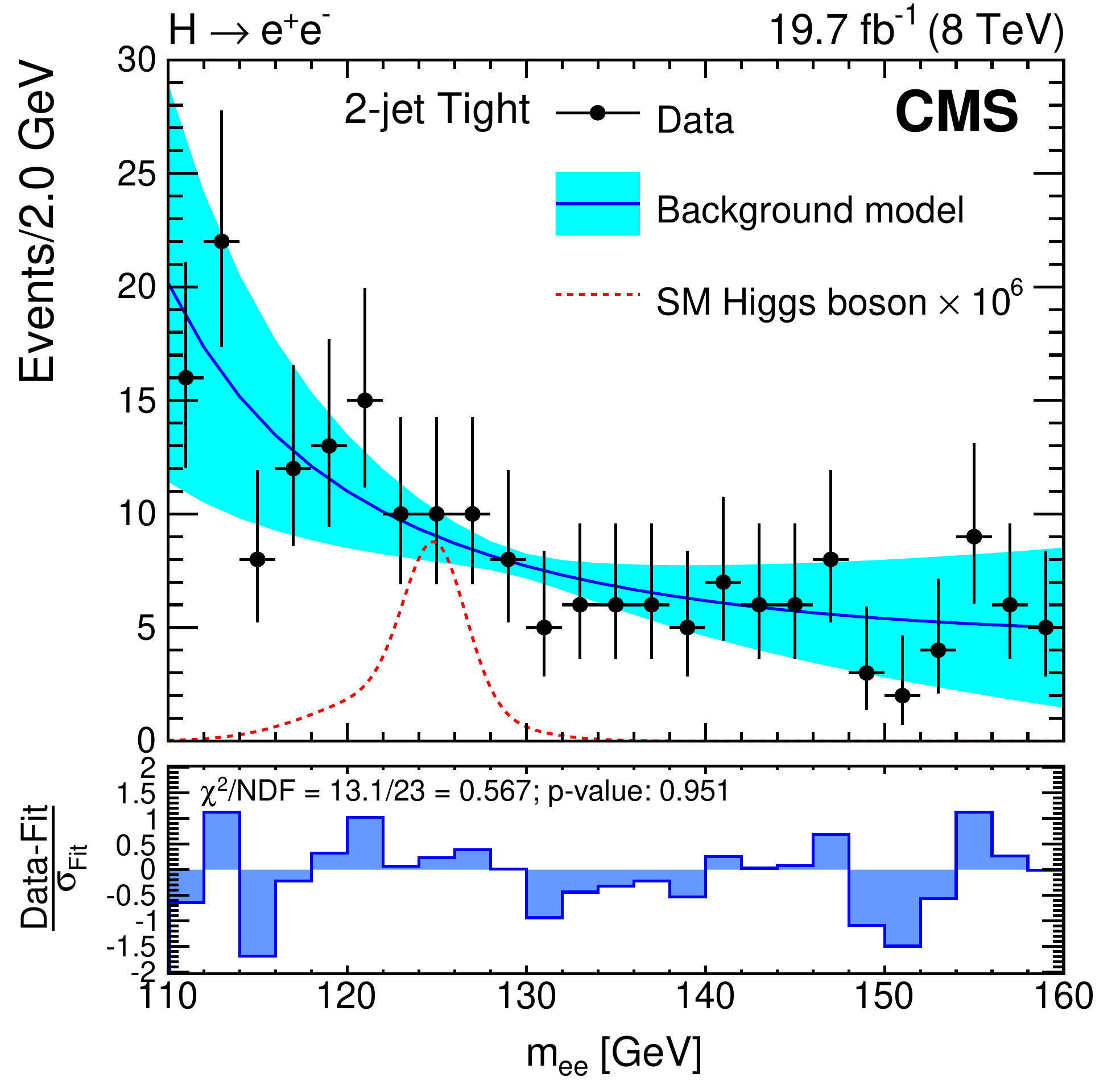}
    \caption{ The dielectron invariant mass at 8\TeV and the background model are shown
      for the 0,1-jet BB (\cmsLeft) and 2-jet Tight (\cmsRight) categories. A best fit of the background model
      (see Section~\ref{sec:results}) is shown
      by a solid line, while its fit uncertainty is represented by a lighter band.
      The dotted line illustrates the expected SM Higgs boson signal enhanced by a factor of $10^6$,
      for $\mh=125\GeV$.
      The lower histograms show the residual for each bin (Data-Fit) normalized by
      the Poisson statistical uncertainty of the background model ($\sigma_\mathrm{Fit}$).
      Also given are the sum of squares
      of the normalized residuals ($\chi^2$) divided by the number of degrees of freedom (NDF) and
      the corresponding $p$-value assuming the sum follows the $\chi^2$ distribution.}
\label{fig:eemassspectrum}

\end{figure}

Results are extracted from the data for $\mh$ values between 120 and 150\GeV
by fitting the mass spectra of the four categories in the range $110 < \Mee <160\GeV$.
The parameterizations used for the signal and background are the same as used in the \mumu search,
a double-Gaussian function and Eq.~(\ref{eqn:bkg}), respectively.
Background-only \Mee fits to data are shown in Fig.~\ref{fig:eemassspectrum} for the 0,1-jet~BB
and 2-jet~Tight categories.

Systematic uncertainties are estimated and incorporated into the results using
the same methods as in the \mumu search (see Section~\ref{sec:systUnc}).
Table~\ref{tab:heeSyst} lists the systematic uncertainties in the signal yield.
The pileup modeling, pileup jet rejection, and MC statistics systematic
uncertainties are small and neglected for the \ee search.  The systematic
uncertainties due to the jet energy resolution and absolute jet energy scale
are combined and listed as ``Jet energy scale'' in Table~\ref{tab:heeSyst}.  The
uncertainty related to the choice of background parameterization in terms of
the number of signal events ($N_\text{P}$) is shown in
Table~\ref{tab:nEvtsHee}.  This systematic uncertainty is larger than all of
the others, and removing it would lower the expected limit by 28\%, for
$\mh=125\GeV$.

\begin{table*}[!hbtp]
  \centering
    \topcaption{ \label{tab:heeSyst}
    The relative systematic uncertainty in the \hee{} signal yield is listed for
    each uncertainty source.  Uncertainties are shown for the GF and VBF Higgs
    boson production modes.  The systematic uncertainties vary depending on the
    category and centre-of-mass energy.
     }
\begin{tabular}{ l c c } \hline
Source                                    & GF [\%]        & VBF [\%]   \\ \hline
Higher-order corrections~\cite{LHCHXSWG3} & 8--18           & 1--7        \\
PDF~\cite{LHCHXSWG3}                      & 11             & 5          \\
PS/UE                                     & 6--42           & 3--10       \\
Integrated luminosity~\cite{lumi_2013}    & 2.6            & 2.6        \\
Electron efficiency                       & 2              & 2          \\
Jet energy scale                          & $<$1--11        & 2--3        \\ \hline
\end{tabular}
\end{table*}

No significant excess of events is observed.
Upper limits on $\sigma(8\TeV)  \BF$ and $\BF$ are reported.
The observed 95\% CL upper limit on $\sigma(8\TeV)  \BF$
at 125\GeV
is 0.041\unit{pb} while the background-only expected limit is $0.052^{+0.022}_{-0.015}$\unit{pb}.
Assuming the SM production cross section, this
corresponds to an observed upper limit on $\BF(\hee)$
of 0.0019, which is approximately $3.7\times10^5$ times the SM prediction.
Upper limits on $\sigma(8\TeV)  \BF$
are shown for Higgs boson masses from 120 to 150\GeV at the 95\% CL in Fig.~\ref{fig:xsbrLimits} (
\cmsRight).

\begin{table*}[!hbtp]
  \centering
    \topcaption{ \label{tab:nEvtsHee}
        Details regarding each category of the \hee analysis,
        for $19.7\pm0.5$\fbinv at 8\TeV.
        Each row lists the category name,
        FWHM of the signal peak,
        acceptance times selection efficiency ($A\epsilon$) for GF,
        $A\epsilon$ for VBF,
        expected number of SM signal events in the category times $10^5$ for
        $\mh=125\GeV$ ($N_\mathrm{S}$),
        number of background events within a FWHM-wide window centered on 125\GeV
        estimated by a signal plus background fit to the data ($N_\mathrm{B}$),
        number of observed events within a FWHM-wide window centered
        on 125\GeV ($N_\text{Data}$), systematic uncertainty
        to account for the parameterization of the background ($N_\mathrm{P}$), and
        $N_\mathrm{P}$ divided by the statistical uncertainty on the fitted
        number of signal events ($N_\mathrm{P}/\sigma_\text{Stat}$).
        The expected number of SM signal events is
        $N_\mathrm{S} = \mathcal{L}\times (\sigma  \BF  A  \epsilon)_\mathrm{GF}+\mathcal{L}\times (\sigma  \BF  A  \epsilon)_\mathrm{VBF}$,
        where $\mathcal{L}$ is the integrated luminosity and $\sigma \BF$ is the SM cross section
        times branching fraction.
    }
    \begin{tabular}{ld{1.1}d{2.2}d{2.2}d{1.1}d{4.1}d{5.1}d{2.2}d{2.0}} \hline
     & \mc{FWHM}
     & \multicolumn{2}{c}{$A\epsilon$ [\%]} & & & &
     & \mc{$N_\mathrm{P}/\sigma_\text{Stat}$} \\ \cline{3-4}
     Category & \mc{[\GeVns{}]}
        & \mc{GF} & \multicolumn{1}{c}{VBF}
        & \mc{$N_\mathrm{S}\times10^5$}
        & \mc{$N_\mathrm{B}$}
        & \mc{$N_\mathrm{Data}$}
        & \mc{$N_\mathrm{P}$}
        & \mc{[\%]} \\ \hline
0,1-jet BB        & 4.0    & 27.5    &  16.7   &  56.1  &  5208.9      &  5163         & 75.0  &   61 \\
0,1-jet Not BB    & 7.1    & 17.0    &  9.7    &  34.6  &  8675.0      &  8748         & 308.7 &  174 \\
2-jet Tight       & 3.8    & 0.5     &  10.7   &  2.6   &  17.7        &    22         & 19.5  &   71 \\
2-jet Loose       & 4.7    & 1.0     &  7.3    &  3.1   &  79.5        &    84         & 43.2  &   88 \\
Sum of categories & \mcd   & 46.0    &  44.4   &  96.4  &  13981.1     & 14017         & \mcd  & \mcd \\
     \hline
    \end{tabular}
\end{table*}

 \section{Summary}
\label{sec:summary}

Results are presented from a search for a SM-like Higgs boson decaying to
\mumu and for the first time to \ee.
For the search in \mumu, the analyzed CMS data correspond to integrated
luminosities of $5.0\pm0.1$\fbinv collected at
7\TeV and $19.7\pm0.5$\fbinv collected
at 8\TeV, while only the 8\TeV data are used for the search in the \ee channel.
The Higgs boson signal is sought as a narrow peak in the
dilepton invariant mass spectrum on top of a smoothly falling background
dominated by the Drell--Yan, \ttbar, and vector boson pair-production processes.
Events are split into categories corresponding to different
production topologies and dilepton invariant mass resolutions.
The signal strength is then extracted using a simultaneous fit to
the dilepton invariant mass spectra in all of the categories.

No significant \hmm signal is observed. Upper limits are set on the signal strength
at the 95\% CL.
 Results are presented for Higgs boson masses
between 120 and 150\GeV.
The combined observed limit on the signal strength,
for a Higgs boson with a mass of 125\GeV,
is 7.4, while
the expected limit is $6.5^{+2.8}_{-1.9}$.
Assuming the SM production cross section, this corresponds to an upper
limit of 0.0016 on $\BF(\hmm)$.
For a Higgs boson mass of 125\GeV, the best fit signal strength
is $0.8^{+3.5}_{-3.4}$.

In the \hee channel, SM Higgs boson decays are far too rare to detect, and no signal is observed.
For a Higgs boson mass of 125\GeV, a 95\% CL upper limit
of 0.041\unit{pb} is set on $\sigma  \BF(\hee)$ at 8\TeV.
Assuming the SM production cross section, this corresponds to an upper limit
on $\BF(\hee)$ of 0.0019, which is approximately $3.7\times10^5$
times the SM prediction.
For comparison, the \hmm observed 95\% CL upper limit on
$\sigma  \BF(\hmm)$ is 0.033\unit{pb} (using only 8\TeV data),
which is 7.0 times the expected SM Higgs boson cross section.

These results, together with recent evidence for the 125\GeV boson's coupling to
$\tau$-leptons~\cite{cmsHtautau} with a larger \BF
 consistent with the SM value of $0.0632 \pm 0.0036$~\cite{Denner_2011mq},
confirm the SM prediction that the leptonic couplings of the new boson are
not flavour-universal.

\begin{acknowledgments}
We congratulate our colleagues in the CERN accelerator departments for the excellent performance of the LHC and thank the technical and administrative staffs at CERN and at other CMS institutes for their contributions to the success of the CMS effort. In addition, we gratefully acknowledge the computing centres and personnel of the Worldwide LHC Computing Grid for delivering so effectively the computing infrastructure essential to our analyses. Finally, we acknowledge the enduring support for the construction and operation of the LHC and the CMS detector provided by the following funding agencies: BMWFW and FWF (Austria); FNRS and FWO (Belgium); CNPq, CAPES, FAPERJ, and FAPESP (Brazil); MES (Bulgaria); CERN; CAS, MoST, and NSFC (China); COLCIENCIAS (Colombia); MSES and CSF (Croatia); RPF (Cyprus); MoER, ERC IUT and ERDF (Estonia); Academy of Finland, MEC, and HIP (Finland); CEA and CNRS/IN2P3 (France); BMBF, DFG, and HGF (Germany); GSRT (Greece); OTKA and NIH (Hungary); DAE and DST (India); IPM (Iran); SFI (Ireland); INFN (Italy); NRF and WCU (Republic of Korea); LAS (Lithuania); MOE and UM (Malaysia); CINVESTAV, CONACYT, SEP, and UASLP-FAI (Mexico); MBIE (New Zealand); PAEC (Pakistan); MSHE and NSC (Poland); FCT (Portugal); JINR (Dubna); MON, RosAtom, RAS and RFBR (Russia); MESTD (Serbia); SEIDI and CPAN (Spain); Swiss Funding Agencies (Switzerland); MST (Taipei); ThEPCenter, IPST, STAR and NSTDA (Thailand); TUBITAK and TAEK (Turkey); NASU and SFFR (Ukraine); STFC (United Kingdom); DOE and NSF (USA).

Individuals have received support from the Marie-Curie programme and the European Research Council and EPLANET (European Union); the Leventis Foundation; the A. P. Sloan Foundation; the Alexander von Humboldt Foundation; the Belgian Federal Science Policy Office; the Fonds pour la Formation \`a la Recherche dans l'Industrie et dans l'Agriculture (FRIA-Belgium); the Agentschap voor Innovatie door Wetenschap en Technologie (IWT-Belgium); the Ministry of Education, Youth and Sports (MEYS) of the Czech Republic; the Council of Science and Industrial Research, India; the HOMING PLUS programme of Foundation for Polish Science, cofinanced from European Union, Regional Development Fund; the Compagnia di San Paolo (Torino); the Consorzio per la Fisica (Trieste); MIUR project 20108T4XTM (Italy); the Thalis and Aristeia programmes cofinanced by EU-ESF and the Greek NSRF; and the National Priorities Research Program by Qatar National Research Fund; and the Russian Scientific Fund,
grant N 14-12-00110.
\end{acknowledgments}

 \bibliography{auto_generated}   
\cleardoublepage \appendix\section{The CMS Collaboration \label{app:collab}}\begin{sloppypar}\hyphenpenalty=5000\widowpenalty=500\clubpenalty=5000\textbf{Yerevan Physics Institute,  Yerevan,  Armenia}\\*[0pt]
V.~Khachatryan, A.M.~Sirunyan, A.~Tumasyan
\vskip\cmsinstskip
\textbf{Institut f\"{u}r Hochenergiephysik der OeAW,  Wien,  Austria}\\*[0pt]
W.~Adam, T.~Bergauer, M.~Dragicevic, J.~Er\"{o}, C.~Fabjan\cmsAuthorMark{1}, M.~Friedl, R.~Fr\"{u}hwirth\cmsAuthorMark{1}, V.M.~Ghete, C.~Hartl, N.~H\"{o}rmann, J.~Hrubec, M.~Jeitler\cmsAuthorMark{1}, W.~Kiesenhofer, V.~Kn\"{u}nz, M.~Krammer\cmsAuthorMark{1}, I.~Kr\"{a}tschmer, D.~Liko, I.~Mikulec, D.~Rabady\cmsAuthorMark{2}, B.~Rahbaran, H.~Rohringer, R.~Sch\"{o}fbeck, J.~Strauss, A.~Taurok, W.~Treberer-Treberspurg, W.~Waltenberger, C.-E.~Wulz\cmsAuthorMark{1}
% ...
\vskip\cmsinstskip
% ...
1:~~Also at Vienna University of Technology, Vienna, Austria\\
2:~~Also at CERN, European Organization for Nuclear Research, Geneva, Switzerland\\
% ...

\end{sloppypar}
\end{document}